\documentclass[12pt]{article}
\pdfoutput=1
\usepackage{graphicx}
\usepackage{url}
\def\beq{\begin{equation}}
\def\eeq#1{\label{#1}\end{equation}}
\def\eeqn{\end{equation}}

\def\beqa{\begin{eqnarray}}
\def\eeqa#1{\label{#1}\end{eqnarray}}
\def\eeqan{\end{eqnarray}}

\let\bar=\overbar

\def\Dslash{\not{\hbox{\kern-4pt $D$}}}
\def\dslash{\not{\hbox{\kern-2pt $\del$}}}

\def\msb{{\bar{\ssstyle M \kern -1pt S}}}

\def\Title#1{\begin{center} {\Large {\bf #1} } \end{center}}

\begin{document}

\Title{$CP$ violation and mixing in the charm sector at Belle, and current HFAG averages}
\bigskip\bigskip
\begin{raggedright}  
{\it Byeong Rok Ko (for the Belle collaboration)\\
Department of Physics\\
College of Science, Korea University\\
SEOUL, 136-713 Republic of Korea}
\bigskip\bigskip
\end{raggedright}
\section{Introduction}
Current HFAG averages of the mixing and $CP$ violation ($CPV$)
parameters in the $D^0$ meson system is to be reviewed. We present
recent Belle measurements of the mixing parameters $x$ and $y$ using
$D^0\rightarrow K^0_S\pi^+\pi^-$, and $y_{CP}$ and $A_\Gamma$ from
$D^0\rightarrow h^+h^-$, $D^0\rightarrow K^-\pi^+$, where $h$ denotes
$K$ or $\pi$. Belle measurements of direct $CPV$ in $D^+\rightarrow
K^0_S\pi^+$, $D^0\rightarrow h^+h^-$, and $D^+\rightarrow K^0_SK^+$
decays together with $\Delta A^{hh}_{CP}$ are to be presented.
\section{Current HFAG averages~\cite{HFAG}} 
$D^0-\bar{D}^0$ mixing occurs since the mass eigenstates $D_1$ and
$D_2$ are different from the weak eigenstates $D^0$ and
$\bar{D}^0$. Assuming $CPT$ is conserved, the mass eigenstates can be
written in terms of the weak eigenstates by
$|D_1\rangle=p|D^0\rangle-q|\bar{D}^0\rangle$ and
$|D_2\rangle=p|D^0\rangle+q|\bar{D}^0\rangle$, where
$|p|^2+|q|^2=1$. Thus without $CPV$ $D_1$ and $D_2$ are $CP$-even and
$CP$-odd, respectively, with the HFAG convention
$CP|D^0\rangle=-|\bar{D}^0\rangle$ and
$CP|\bar{D}^0\rangle=-|D^0\rangle$. The mixing parameters, $x$ and $y$
can be expressed by the difference of masses and widths between the
two mass eigenstates, $x=(m_1-m_2)/\Gamma$ and
$y=(\Gamma_1-\Gamma_2)/2\Gamma$, where
$\Gamma=(\Gamma_1+\Gamma_2)/2$. $CPV$ parameters are $|q/p|$ and
$\phi=Arg(q/p)$, where the former and the latter are responsible for
$CPV$ in mixing and that in interference between the decays with and
without mixing, respectively. Without direct $CPV$, alternative mixing
and $CPV$ parameters are $x_{12}=2|M_{12}|/\Gamma$,
$y_{12}=|\Gamma_{12}|/\Gamma$, and
$\phi_{12}=Arg(M_{12}/\Gamma_{12})$, where $M_{12}$ and $\Gamma_{12}$
are off-diagonal elements of the $D^0-\bar{D}^0$ mass and decay
matrices which are responsible for the mixing. Without $CPV$ $x_{12}$,
$y_{12}$, and $\phi_{12}$ become $x$, $y$, and zero,
respectively. Current HFAG averages of the mixing parameters $x$, $y$,
$x_{12}$, and $y_{12}$ rule out the no-mixing hypothesis with more than
10$\sigma$ significance, but they are difficult to be interpreted with
the standard model (SM) due to the final state
interactions~\cite{MIXING_FSI}. $CPV$ parameters, $|q/p|$, $\phi$, and
$\phi_{12}$ show no indirect $CPV$ at present. Since both $x$ and $y$
favor positive values, the $CP$-even state in the $D^0$ system is
heavier (unlike $K^0$ system) and shorter-lived (like $K^0$ system).
\section{$CPV$ and mixing in the charm sector at Belle}
\subsection{Mixing ($x$, $y$ from $D^0\rightarrow K^0_S\pi^+\pi^-$ and
  $y_{CP}$, $A_{\Gamma}$ from $D^0\rightarrow h^+h^-$ and $D^0\rightarrow K^-\pi^+$)}
\begin{figure}[htbp]
\begin{center}
  \includegraphics[width=1.0\textwidth]{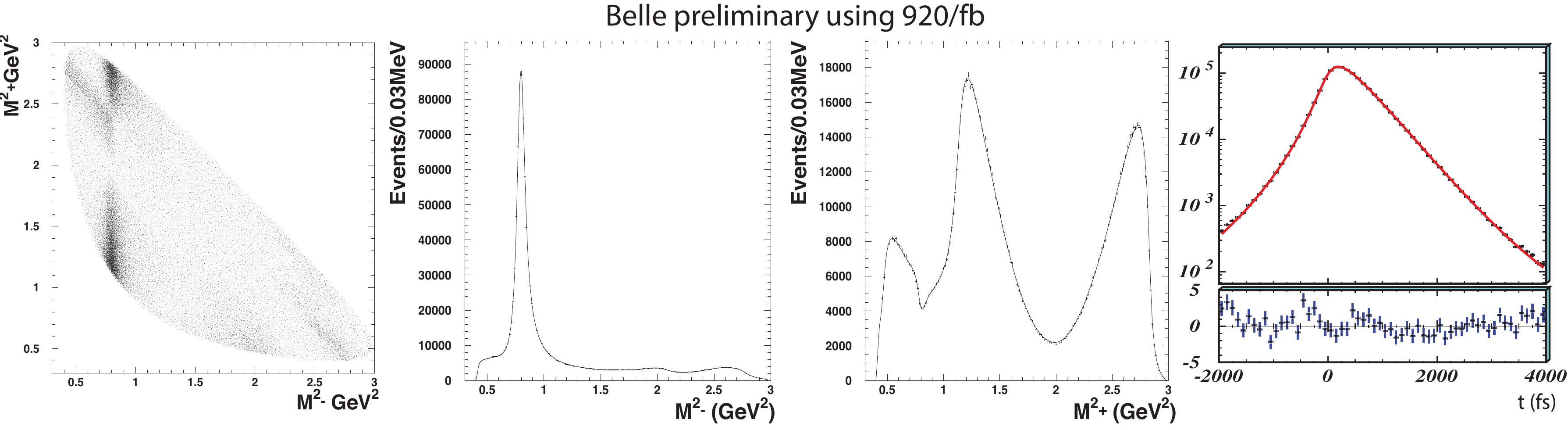}
\caption{From left, 2-D Dalitz plot, fit projections onto $m^2_-$,
  $m^2_+$, and proper decay time, respectively.}
\label{FIG:XY}
\end{center}
\end{figure}

Time dependent decay matrix element of $D^0\rightarrow
K^0_S\pi^+\pi^-$ is expressed with
\begin{equation}
  \mathcal{M}(m^2_-,m^2_+,t)=\mathcal{A}(m^2_-,m^2_+)\frac{e_1(t)+e_2(t)}{2}-\frac{q}{p}\bar{\mathcal{A}}(m^2_+,m^2_-)\frac{e_1(t)-e_2(t)}{2},
  \label{EQ:ME}
\end{equation}
where $t$ is proper decay time, $m^2_{\mp}=m^2(K^0_S\pi^{\mp})$,
$e_j(t)=e^{-t(\Gamma_j/2 + im_j)}$, and
$\mathcal{A}(\bar{\mathcal{A}})$ is the decay amplitude of
$D^0(\bar{D}^0)$. Thus the mixing parameters, $x$ and $y$ can be
extracted with time dependent Dalitz analysis of the decay rate,
$\mathcal{M}^2$. The best fit model for $\mathcal{A}(m^2_-,m^2_+)$ is
found to be a sum of Breit-Wigner for P-, D-wave resonances,
K-matrix~\cite{KMATRIX} and LASS~\cite{LASS} models for $\pi\pi$ and
$K\pi$ S-wave states, respectively, without non-resonant
decay. Figure~\ref{FIG:XY} shows the 2-D Dalitz plot and fit results
as $m^2_{\mp}$ and decay time projections. The results from the fit
under $CP$ conservation, $q/p=1$ and $\mathcal{A}=\bar{\mathcal{A}}$,
are $x=(0.56\pm0.19^{+0.03+0.06}_{-0.09-0.09})\%$,
$y=(0.30\pm0.15^{+0.04+0.03}_{-0.05-0.06})\%$, and
$\tau_{D^0}=(410.3\pm0.4)$ fs, where $x$ and $y$ are the most
sensitive to date and $\tau_{D^0}$ is consistent with world
average~\cite{PDG2012}.

\begin{figure}[htbp]
\begin{center}
  \includegraphics[width=1.0\textwidth]{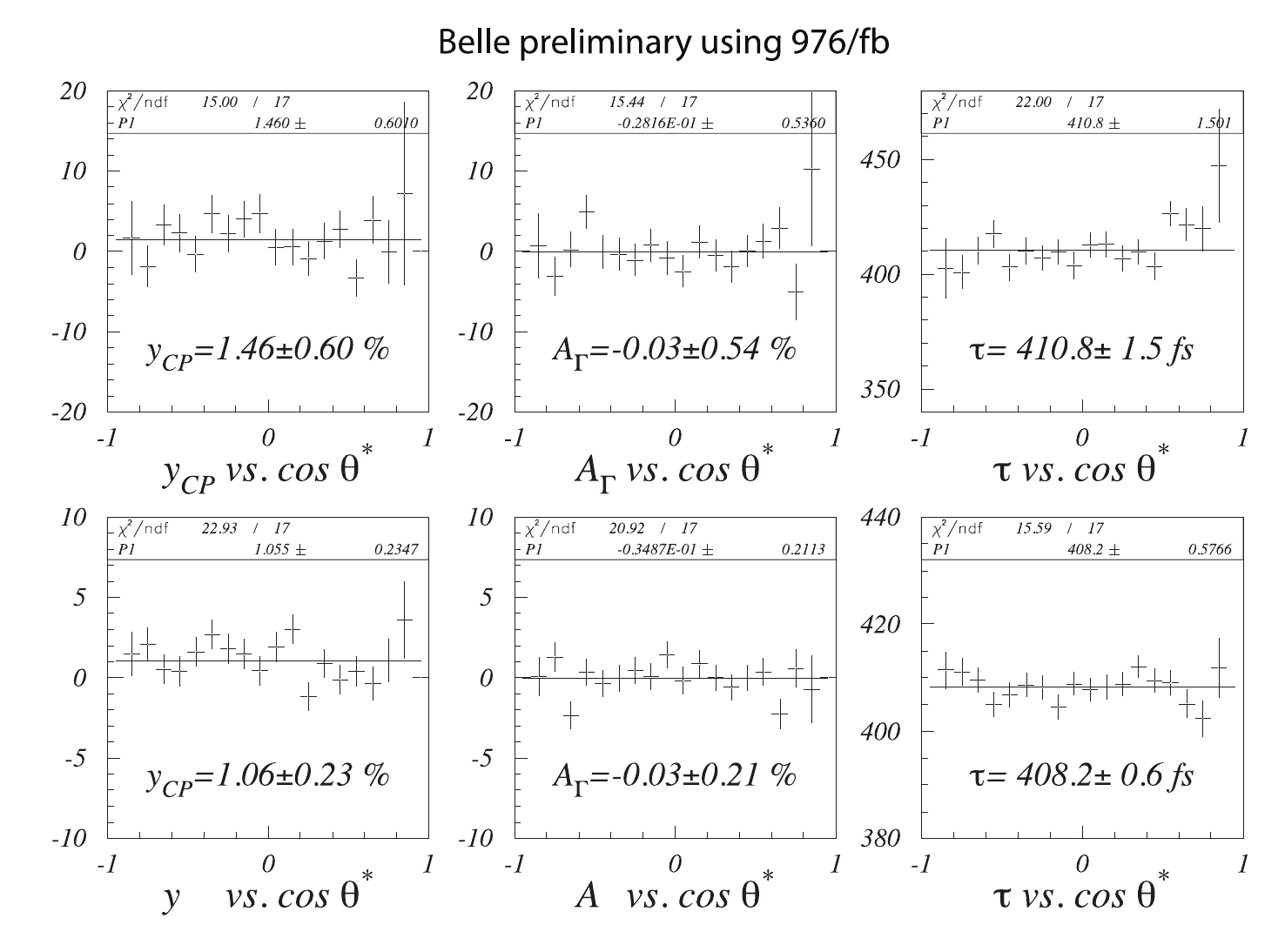}
\caption{$y_{CP}$, $A_\Gamma$, and $\tau_{D^0\rightarrow K^-\pi^+}$ as
  a function of the $\cos\theta^{*}$. Top(bottom) three plots are
  obtained with 3-layer(4-layer) silicon detector.}
\label{FIG:YCP}
\end{center}
\end{figure}

The width asymmetry between neutral $D~CP$-even and $CP$-odd
eigenstates is referred to as $y_{CP}$ and the experimental observable
for $y_{CP}$ is the lifetime difference between $CP$-even
($D^0\rightarrow h^+h^-$) and $CP$-mixed ($D^0\rightarrow K^-\pi^+$)
states as shown in Eq.~(\ref{EQ:YCP}). The $y_{CP}$ could be different
from $y$ if $CP$ is violated in charm decays.
\begin{equation}
  y_{CP}=\frac{\Gamma(CP{\rm-even})-\Gamma(CP{\rm-odd})}{\Gamma(CP{\rm-even})+\Gamma(CP{\rm-odd})}=\frac{\tau_{D^0\rightarrow K^-\pi^+}}{\tau_{D^0\rightarrow h^+h^-}}-1\simeq y.
  \label{EQ:YCP}
\end{equation}
The width asymmetry between two $CP$ conjugate modes provides the
$CPV$ parameter, $A_\Gamma$ which can be measured from lifetime
difference between the two $CP$ conjugate decays. Figure~\ref{FIG:YCP}
show the results for $y_{CP}$, $A_\Gamma$, and $\tau_{D^0\rightarrow
K^-\pi^+}$ as a function of the $\cos\theta^{*}$, where $\theta^{*}$
is polar angle of $D^0$ at the center-of-mass system (c.m.s.). The
averages are $y_{CP}=(1.11\pm0.22\pm0.11)\%$,
$A_\Gamma=(-0.03\pm0.20\pm0.08)\%$, and $\tau_{D^0\rightarrow
K^-\pi^+}=(408.56\pm0.54)$ fs, where the last is consistent with world
average~\cite{PDG2012}. Thus we observe $y_{CP}$ with 4.5$\sigma$
significance and find no $CPV$.
\subsection{Direct $CPV$ ($A_{CP}$ in $D^+\rightarrow
  K^0_S\pi^+$, $D^0\rightarrow h^+h^-$, $D^+\rightarrow K^0_S K^+$,
  and $\Delta A^{hh}_{CP}$)}
The $D^+\rightarrow K^0_S\pi^+$ final state is a coherent sum of
Cabibbo-favored and doubly Cabibbo-suppressed decays where no SM $CPV$
in charm decay is expected, while $(-0.332\pm0.006)\%$~\cite{PDG2012}
of $A^{\bar{K}^0}_{CP}$~\footnote{Here, $A^{\bar{K}^0}_{CP}$ denotes
  $CPV$ due to $K^0-\bar{K}^0$ mixing.} is expected. Using $\sim$1.74M
reconstructed $D^+\rightarrow K^0_S\pi^+$ decays, the $A_{CP}$ is
measured in bins of $|\cos\theta^{\rm c.m.s.}_{D^+}|$ and the average
of $A^{D^+\rightarrow K^0_S\pi^+}_{CP}$ is
$(-0.363\pm0.094\pm0.067)\%$ which shows 3.2$\sigma$ deviations from
zero. This is the first evidence for $CPV$ in charm decays from a
single decay mode while the measured asymmetry is consistent with the
$A^{\bar{K}^0}_{CP}$. After subtracting experiment dependent
$A^{\bar{K}^0}_{CP}$~\cite{GROSSMAN_NIR}, the $CPV$ in charm decay,
$A^{D^+\rightarrow\bar{K}^0\pi^+}_{CP}$~\footnote{We neglect doubly
  Cabibbo-suppressed decay, $D^+\rightarrow K^0\pi^+$.}, is measured
to be $(-0.024\pm0.094\pm0.067)\%$~\cite{KSPIPRL}.
\begin{figure}[htbp]
\begin{center}
\mbox{\includegraphics[width=0.6\textwidth]{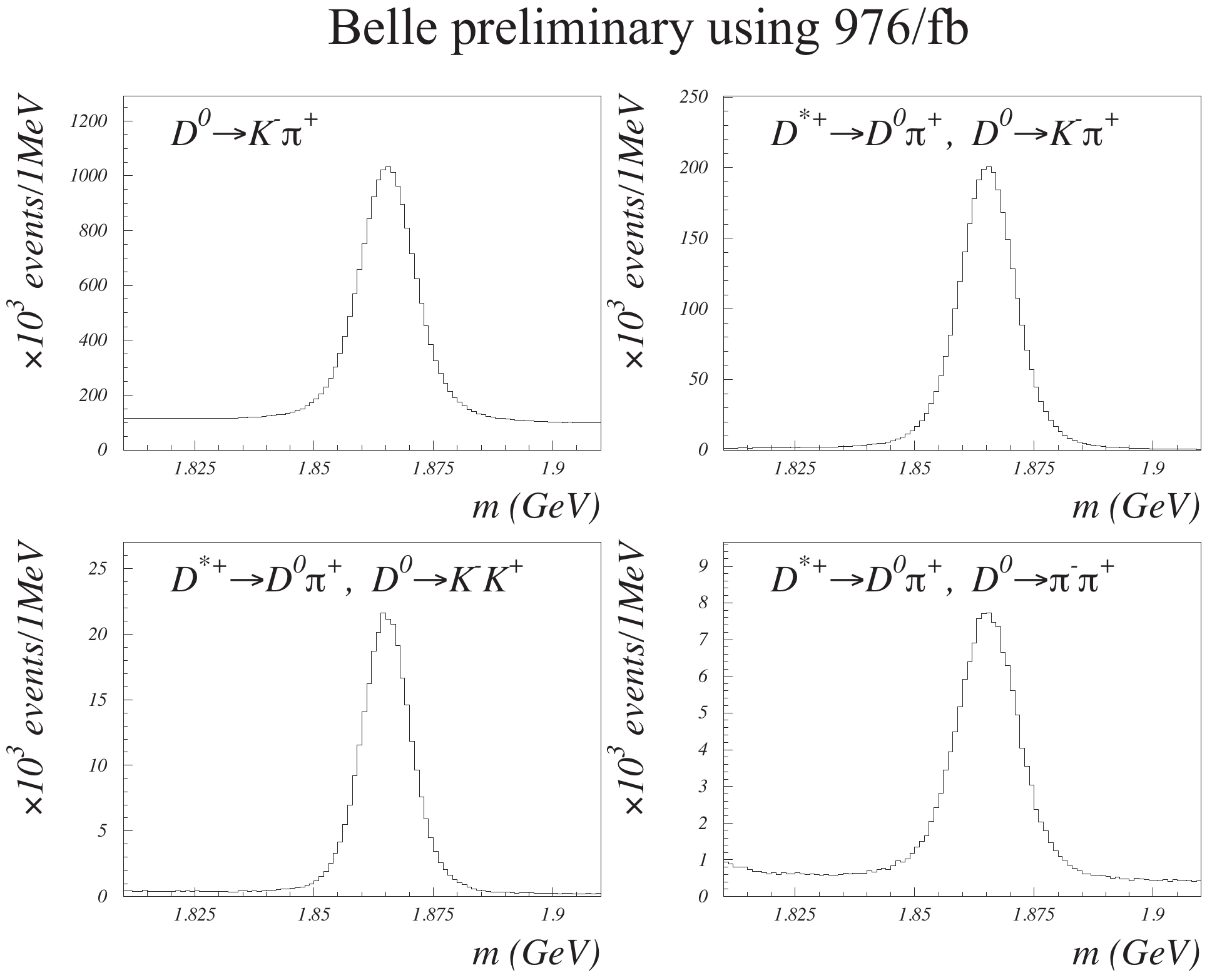}}
\mbox{\includegraphics[width=0.6\textwidth]{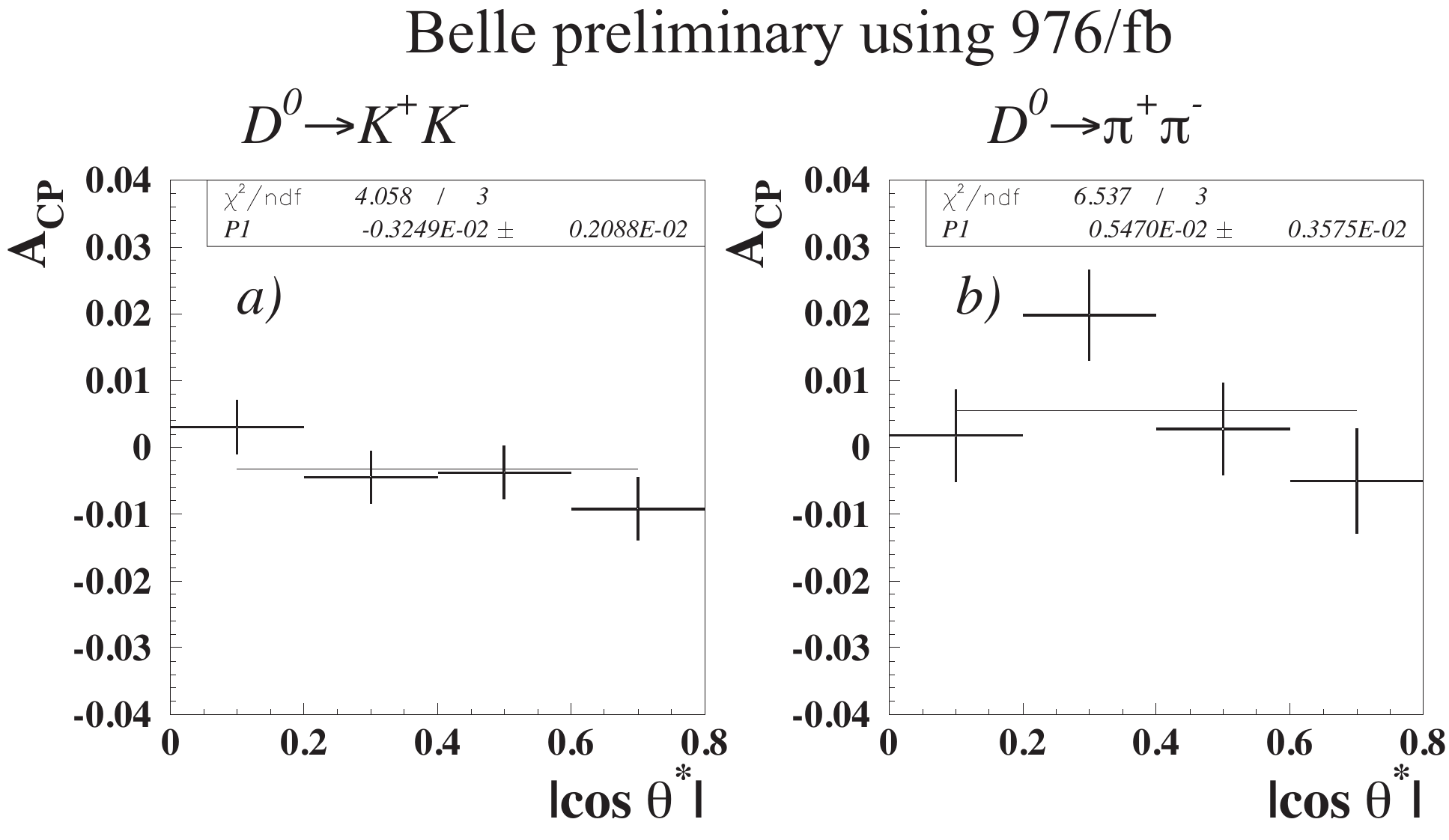}}
\caption{Top four plots show reconstructed signal distributions
  described in the text and bottom two plots show preliminary results
  of $A_{CP}$ as a function of the polar angle of $D^{*+}$ momentum at
  the c.m.s.}
\label{FIG:ACPD0hh}
\end{center}
\end{figure}

The $D^0\rightarrow h^+h^-$ final states are singly Cabibbo-suppressed
(SCS) decays in which both direct and indirect $CPV$ are
expected in the SM~\cite{SMCP}, while the $CP$ asymmetry difference
between the two decays, $\Delta
A^{hh}_{CP}=A^{KK}_{CP}-A^{\pi\pi}_{CP}$ reveals approximately direct
$CPV$ with the universality of indirect $CPV$ in
charm decays~\cite{YAY}. Figure~\ref{FIG:ACPD0hh} shows reconstructed
signal distributions showing 14.7M $D^0\rightarrow K^-\pi^+$, 3.1M
$D^{*+}$ tagged $D^0\rightarrow K^-\pi^+$, 282k $D^{*+}$ tagged
$D^0\rightarrow K^+K^-$, and 123k $D^{*+}$ tagged
$D^0\rightarrow\pi^+\pi^-$, respectively, and the measured $A_{CP}$ in
bins of $|\cos\theta^{*}_{D^{*+}}|$. From the bottom plots in
Fig.~\ref{FIG:ACPD0hh}, we obtain
$A^{KK}_{CP}=(-0.32\pm0.21\pm0.09)\%$ and
$A^{\pi\pi}_{CP}=(+0.55\pm0.36\pm0.09)\%$ where the former shows the
best sensitivity to date. From the two measurements, we obtain $\Delta
A^{hh}_{CP}=(-0.87\pm0.41\pm0.06)\%$ which shows 2.1$\sigma$
deviations from zero and supports recent LHCb~\cite{ACP_LHCB} and
CDF~\cite{ACP_CDF} measurements. By combining LHCb, CDF, and Belle
results, the average of $\Delta A^{hh}_{CP}$ becomes
$(-0.74\pm0.15)\%$.

\begin{figure}[htbp]
\begin{center}
\mbox{\includegraphics[width=0.5\textwidth]{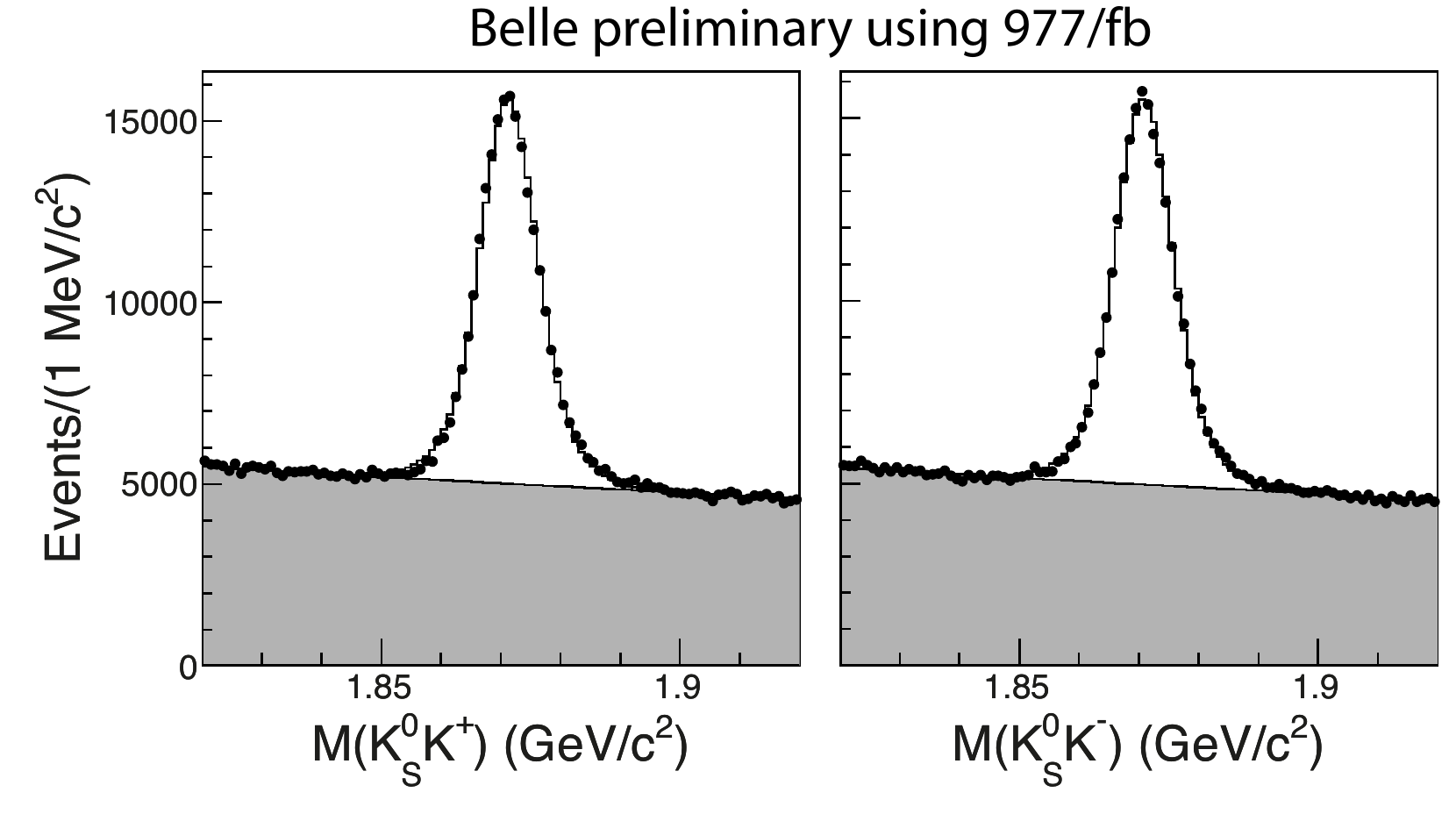}}
\mbox{\includegraphics[width=0.45\textwidth]{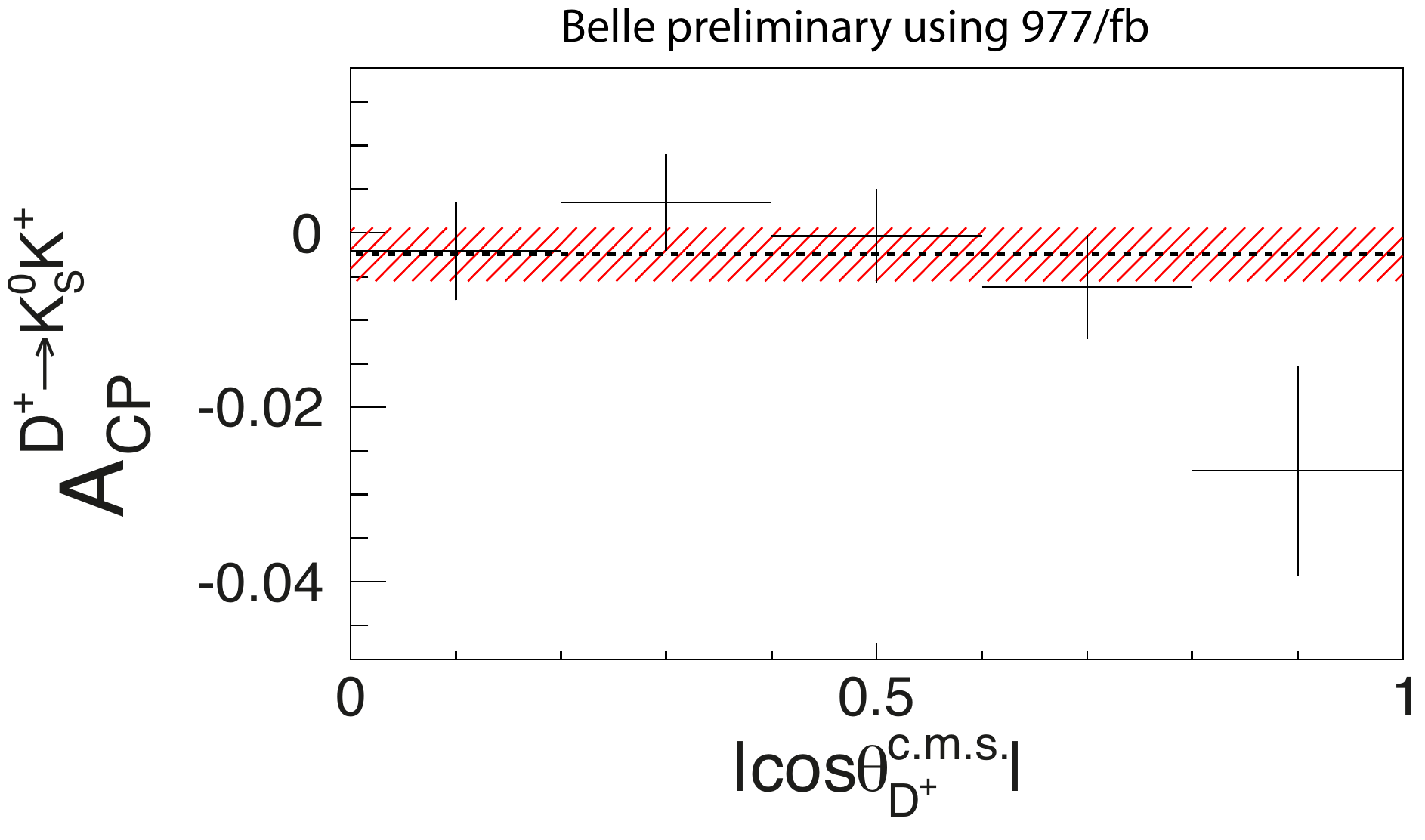}}
\caption{Left two plots show $M(K^0_SK^+)$ and $M(K^0_SK^-)$
  distributions, respectively, and right plot shows preliminary result
  of $A_{CP}$ as a function of the polar angle of $D^{+}$ momentum at
  the c.m.s.}
\label{FIG:ACPKSK}
\end{center}
\end{figure}

The $D^+$ decaying to the final state $K^0_S K^+$ proceeds from
$D^+\rightarrow\bar{K}^0K^+$ decay which is SCS, where direct $CPV$ is
predicted to occur~\cite{SMCP}. The decay $D^+\rightarrow\bar{K}^0
K^+$ shares the same decay diagrams with $D^0\rightarrow K^+K^-$ by
exchanging the spectator quarks, $d\leftrightarrow u$. Therefore,
neglecting the helicity and color suppressed contributions in
$D^+\rightarrow\bar{K}^0 K^+$ and $D^0\rightarrow K^+K^-$ decays, the
direct $CPV$ in the two decays is expected to be effectively the
same. Thus, as a complementary test of the current $\Delta
A^{hh}_{CP}$ measurement, the precise measurement of $A_{CP}$ in
$D^+\rightarrow\bar{K}^0 K^+$ helps to pin down the origin of $\Delta
A^{hh}_{CP}$~\cite{BBACPKSK}. Figure~\ref{FIG:ACPKSK} shows invariant
masses of $D^{\pm}\rightarrow K^0_S K^{\pm}$ together with the fits
that result in $\sim$277k reconstructed decays and the measured
$A_{CP}$ in bins of $|\cos\theta^{\rm c.m.s.}_{D^+}|$. From the right
plot in Fig.~\ref{FIG:ACPKSK}, we obtain $A^{D^+\rightarrow K^0_S
  K^+}_{CP}=(-0.246\pm0.275\pm0.135)\%$. After subtracting experiment
dependent $A^{\bar{K}^0}_{CP}$~\cite{GROSSMAN_NIR}, the $CPV$ in charm
decay, $A^{D^+\rightarrow\bar{K}^0 K^+}_{CP}$, is measured to be
$(+0.082\pm0.275\pm0.135)\%$. The current average of $\Delta
A^{hh}_{CP}$ measurements as well as the Belle preliminary result of
$A^{KK}_{CP}$ favor a negative value. Our result, on the other hand,
does not show this tendency for $D^+\rightarrow\bar{K}^0K^+$ decays,
albeit with a significant statistical uncertainty.
\section{Summary}
In summary, we review the current HFAG averages of $CPV$ and mixing
parameters in the charm sector and recent relevant measurements from
Belle. Belle measurements of mixing and $CPV$ parameters are
consistent with current HFAG averages. No direct $CPV$ in charm decays
has been observed from Belle to date.


\begin{thebibliography}{99}

\bibitem{HFAG} 
Y. Amhis {\it et al.} (Heavy Flavor Averaging Group),
\url{arXiv:1207.1158v1 [hep-ex]} and online update at
\url{http://www.slac.stanford.edu/xorg/hfag/}.

\bibitem{MIXING_FSI}
Z. -Z. Xing, Phys. Rev. D {\bf 55}, 196 (1997); S Bianco,
F. L. Fabbri, D. Benson, and I. Bigi, Riv. Nuovo Cim. {\bf 26N7}, 1
(2003); G. Burdman and I. Shipsey, Ann. Rev. Nucl. Part. Sci. {\bf
  53}, 431 (2003).

\bibitem{KMATRIX}
V. V. Anisovich and A. V. Sarantsev, Eur. Phys. J. A {\bf 16}, 229 (2003).

\bibitem{LASS}
D. Aston {\it et al.} (LASS), Nucl. Phys. B {\bf 296}, 493 (1988).

\bibitem{PDG2012}
J. Beringer {\it et al.} (Particle Data Group), Phys. Rev. D {\bf 86},
010001 (2012).

\bibitem{GROSSMAN_NIR} 
Y. Grossman and Y. Nir, JHEP {\bf 2012}, 04 (2012) 002.

\bibitem{KSPIPRL} 
B. R. Ko {\it et al.} (Belle Collab.), Phys. Rev. Lett. {\bf 109}, 021601 (2012), Erratum-ibid. {\bf 109}, 119903 (2012).

\bibitem{SMCP}
F. Buccella {\it et al.}, Phys. Rev. D {\bf 51}, 3478 (1995).

\bibitem{YAY}
Y. Grossman, A. L. Kagan, and Y. Nir, Phys. Rev. D {\bf 75}, 036008 (2007).

\bibitem{ACP_LHCB}
R. Aaij {\it et al.} (LHCb Collaboration), Phys. Rev. Lett. {\bf 108}, 111602 (2012); 

\bibitem{ACP_CDF}
T. Aaltonen {\it et al.} (CDF Collaboration), Phys. Rev. Lett. {\bf 109}, 111801 (2012).

\bibitem{BBACPKSK}
B. Bhattacharya, M. Gronau, and J. L. Rosner, Phys. Rev. D {\bf 85}, 054104 (2012).

\end{thebibliography}
\end{document}